\documentclass[aps,prll,twocolumn,superscriptaddress,amsmath,amssymb]{revtex4}

\usepackage{natbib}
\usepackage{hyperref}
\usepackage{graphicx}
\usepackage{dcolumn}
\usepackage{bm}
\usepackage{hyperref}
\usepackage[mathlines]{lineno}
\usepackage[dvipsnames]{xcolor}

\begin{document}
	
	\title{Magnetic structure and magnetoelastic coupling of GdNiSi$_{3}$ and TbNiSi$_{3}$}
	
	\author{R. Tartaglia}
	\affiliation{``Gleb Wataghin'' Institute of Physics, University of Campinas - UNICAMP, Campinas, S\~ao Paulo 13083-859, Brazil}
	
	\author{F. R. Arantes}
	\affiliation{CCNH, Universidade Federal do ABC (UFABC), 09210-580, Santo Andr\'e, S\~ao Paulo, Brazil}
	
	\author{C. W. Galdino}
	\affiliation{``Gleb Wataghin'' Institute of Physics, University of Campinas - UNICAMP, Campinas, S\~ao Paulo 13083-859, Brazil}
	
	\author{D. Rigitano}
	\affiliation{``Gleb Wataghin'' Institute of Physics, University of Campinas - UNICAMP, Campinas, S\~ao Paulo 13083-859, Brazil}
	
	\author{U. F. Kaneko}
	\affiliation{Brazilian Synchrotron Light Laboratory (LNLS), Brazilian Center for Research in Energy and Materials (CNPEM), Campinas, S\~ao Paulo 13083-970, Brazil}
	
	\author{M. A. Avila} 
	\affiliation{CCNH, Universidade Federal do ABC (UFABC), 09210-580, Santo Andr\'e, S\~ao Paulo, Brazil}
	
	\author{E. Granado} 
	\affiliation{``Gleb Wataghin'' Institute of Physics, University of Campinas - UNICAMP, Campinas, S\~ao Paulo 13083-859, Brazil}
	
	\date{\today}
	
	\begin{abstract}
		
The series of intermetallic compounds $R$NiSi$_3$ ($R$ = rare earth) shows interesting magnetic properties evolving with $R$ and metamagnetic transitions under applied magnetic field for some of the compounds. The microscopic magnetic structures must be determined to rationalize such rich behavior.  Here, resonant x-ray magnetic diffraction experiments are performed on single crystals of GdNiSi$_{3}$ and TbNiSi$_{3}$ at zero field. The primitive magnetic unit cell matches the chemical cell below the N\'eel temperatures $T_{N}$ = 22.2 and 33.2 K, respectively.  The magnetic structure is determined to be the same for both compounds (magnetic space group $Cmmm'$). It features ferromagnetic {\it ac} planes that are stacked in an antiferromagnetic $+-+-$ pattern, with the rare-earth magnetic moments pointing along the $\vec{a}$ direction, which contrasts with the $+--+$ stacking and moment direction along the $\vec{b}$ axis previously reported for YbNiSi$_3$. This indicates a sign reversal of the coupling constant between second-neighbor $R$ planes as $R$ is varied from Gd and Tb to Yb. The long {\it b} lattice parameter of GdNiSi$_{3}$ and TbNiSi$_{3}$ shows a magnetoelastic expansion upon cooling below $T_N$, pointing to the conclusion that the $+-+-$ stacking is stabilized under lattice expansion. A competition between distinct magnetic stacking patterns with similar exchange energies tuned by the size of $R$ sets the stage for the magnetic ground state instability observed along this series.
		
	\end{abstract}
	
	\maketitle
	
	
	\section{Introduction}
	
\begin{figure}
		\includegraphics[width=0.48 \textwidth]{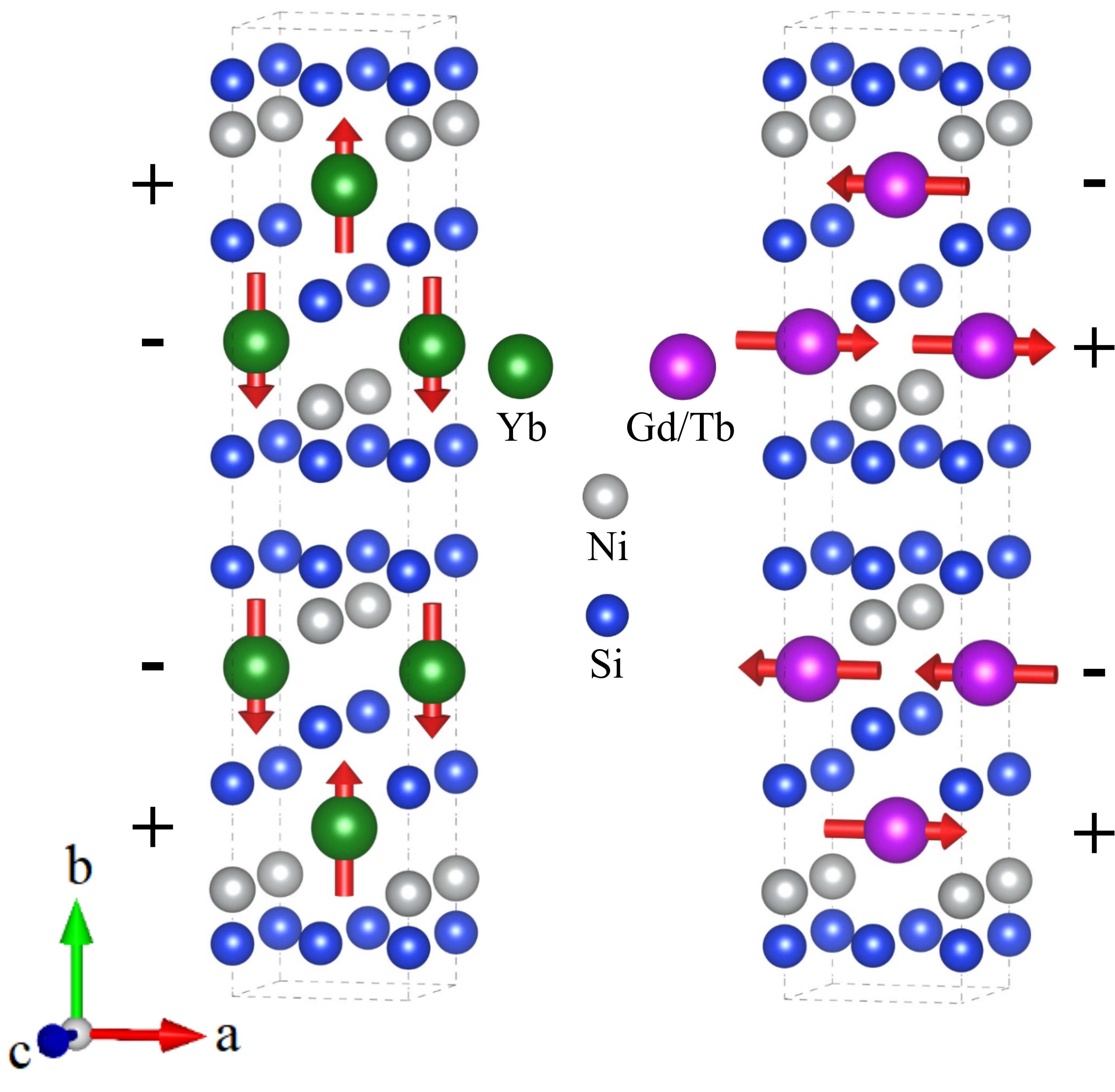}
		\caption{\label{structure} Crystal and magnetic structures of YbNiSi$_3$ \cite{Kobayashi2008}(left), GdNiSi$_3$ and TbNiSi$_3$ (this work, right).}
	\end{figure}

Rare-earth-based materials play a fundamental role in basic and applied condensed matter physics, offering a large amount of fascinating fundamental physical phenomena such as the Kondo effect \cite{Falkowski2016,Magnavita2016}, quantum criticality \cite{Alvarez2004,Honda2012}, unconventional superconductivity \cite{Mathur1998,Wang2012} and others. Combining rare-earth and transition-metal ions usually leads to remarkable magnetic properties such as in Nd$_{2}$Fe$_{14}$B \cite{Min1993,Chen2014} and SmCo$_{5}$ \cite{Kuru2017}, which are widely used as permanent magnets. In fact, such combinations often lead to high magnetic ordering temperatures, characteristic of $3d$ materials, combined to strong magnetic anisotropies, characteristic of the $4f$ moments. On the other hand, there are also cases of intermetallics where the $3d$ element becomes non-magnetic and thus the magnetism is entirely dominated by the $4f$ electrons. One such family of materials is the recently synthesized $R$NiSi$_3$ series (\textit{R} = Y, Gd-Lu) \cite{Avila2004,Arantes2018} with an intricate orthorhombic SmNiGe$_3$-type crystal structure, space group $Cmmm$ [see Fig. \ref{structure}]. The non-magnetic nature of Ni in this structure is demonstrated by the absence of magnetic transitions and local moments in the specific heat and magnetic susceptibility data of YNiSi$_3$ and LuNiSi$_{3}$ \cite{Arantes2018}. 
	
The evolution of the magnetic properties within the $R$NiSi$_3$ series as $R$ is varied from Gd to Yb is very interesting and deserves an individualized description. The temperature- and field-dependent anisotropic bulk magnetization of the compounds with $R$ = Gd-Tm were studied in detail \cite{Aristizabal-Giraldo2015,Arantes2018}. GdNiSi$_3$ ($T_N=22.2$ K) is found to have an easy AFM axis along $\vec{a}$ with a spin-flop transition for a magnetic field along $\vec{a}$ of $B_a=2.7$ T; TbNiSi$_3$ ($T_N=33.2$ K) also shows an AFM axis along $\vec{a}$, displaying a series of step-like metamagnetic transitions with $B_a \gtrsim 4$ T and nearly reaching ferromagnetic saturation for $B_a \sim 6$ T. DyNiSi$_3$, also with an AFM axis along $\vec{a}$, shows two close magnetic transitions, at 22.7 and 23.7 K and metamagnetic transitions towards a ferromagnetic phase for $B_a \sim 2.2-2.7$ T. Remarkably, HoNiSi$_3$ appears to show a component-separated magnetic transition, with two distinct ordering temperatures, one for the AFM axis along $\vec{a}$ ($T_N^{a}=10.0$ K) and another for the AFM axis along $\vec{c}$ ($T_N^{c} \sim 6$ K), also showing field induced metamagnetic transitions towards a partial (i.e., non-saturated) ferromagnetic state for $B_a = B_c \sim 1.2-1.6$ T. ErNiSi$_3$ ($T_N=3.5$ K) shows a dominating AFM direction along $\vec{b}$ with another significant AFM component to the susceptibility along either $\vec{a}$ or $\vec{c}$, displaying metamagnetic transitions towards a partial ferromagnetic state for $B \sim 1-2$ T. TmNiSi$_3$ ($T_N=2.5$ K) shows an AFM direction along $\vec{b}$ and no clear sign of metamagnetic transitions with field at 2~K. Finally, the magnetic end-member YbNiSi$_3$ ($T_N=5.1$ K) was thoroughly studied \cite{Avila2004,Budko2007,Kobayashi2008}, with the AFM axis being along $\vec{b}$. The magnetic structure was solved by neutron diffraction and displays ferromagnetic {\it ac} planes of Yb moments that are stacked antiferromagnetically in a $+--+$ pattern \cite{Kobayashi2008} [see Fig. \ref{structure}(left)]. Bulk measurements indicate a transition to a distinct ordered phase for $B_b = 1.6$ T that is suppressed for a higher field of $B_b \sim 8.5$ T.
	
In order to understand the intriguing evolution of the magnetic properties of the $R$NiSi$_3$ series with $R$, it is necessary to determine the corresponding magnetic structures beyond the already investigated $R$ = Yb end member at zero field \cite{Kobayashi2008}. In this work, we take a step towards this direction by solving the zero-field magnetic structure of the known members with the largest $R$ ionic radius, GdNiSi$_3$ and TbNiSi$_3$, which were found to have similar bulk magnetic properties at small magnetic fields (see above \cite{Arantes2018}). Due to the relatively small crystal sizes and the high neutron absorption coefficient of Gd, resonant x-ray magnetic diffraction was the technique of choice. We find that both the magnetic and chemical structures of these materials adopt the same primitive unit cell and the symmetry-allowed magnetic and charge reflections fall into the same points of the reciprocal space. Polarization analysis was employed to separate these contributions for selected reflections and to determine the magnetic structure of GdNiSi$_3$ and TbNiSi$_3$. It was found to be different to that of YbNiSi$_3$ and our results indicate that the magnetic properties of this family are impacted by an instability of the coupling between magnetic bilayers along the $\vec{b}$-direction.
	
	\section{\label{sec:level1}Experimental Details}
	
Platelet-shaped single crystals of GdNiSi$_{3}$ and TbNiSi$_{3}$ were grown from the melt in Sn flux as described previously \cite{Arantes2018}. 
Sample dimensions for the measured GdNiSi$_{3}$ and TbNiSi$_{3}$ crystals are $1.80\times0.77\times0.08$ mm$^{3}$ and $1.19\times0.75\times0.10$ mm$^{3}$, respectively. The as-grown principal faces correspond the crystallographic {\it ac} plane and rocking curves revealed mosaic widths of 0.10$^{\circ}$ and 0.06$^{\circ}$ full width at half maximum (FWHM) for GdNiSi$_3$ and TbNiSi$_3$, respectively.
	
X-ray diffraction measurements were performed at the x-ray diffraction and spectroscopy (XDS) beamline of the Brazilian Synchrotron Light Laboratory in Campinas, which uses a 4 T superconducting multipolar wiggler source \cite{Lima2016}. The sample was mounted at the cold finger of a closed-cycle He cryostat (base temperature 10 K) with a cylindrical Be window. The cryostat was fixed onto the Eulerian cradle of a commercial 6+2 circle diffractometer, appropriate for single-crystal x-ray diffraction. The energy of the incident photons was selected by a double Si(111) crystal monochromator, with LN$_{2}$ cooling in the first crystal, while the second crystal was bent for sagittal focusing. The beam was vertically focused by a bent Rh-coated mirror placed before the monochromator, which also provided filtering of higher harmonics. Our experiments were performed in the vertical scattering plane, i.e., perpendicular to the linear polarization of the incident photons. A polarimeter stage was mounted downstream a scintillator detector, which enabled selecting either the $\sigma \sigma$' or $\sigma \pi$' polarization channels. For GdNiSi$_3$ data taken near the Gd $L_{II}$ edge, the analyzer material was a highly ordered pyrolytic graphite, yielding 2$\theta$ = 89.44$^{\circ}$ for the 003 reflection at $E=7.924$ keV. For TbNiSi$_{3}$, an Al(111) crystal was used, corresponding to 2$\theta$ = 89.77$^{\circ}$ for $E=7.516$ keV (Tb L$_{II}$ edge). The lattice parameters  were obtained by analyzing the $2 \theta$ angles of the 2 14 0, 0 20 0, and 0 14 $\bar 2$ reflections for GdNiSi$_3$ and 1 19 0, 0 16 0, and 0 14 1 reflections for TbNiSi$_3$.
	
	\section{\label{sec:level1}Results and Analysis}
	
	\subsection{\label{sec:level1}Magnetic Diffraction}
	
	\begin{figure}
		\includegraphics[width=0.48 \textwidth]{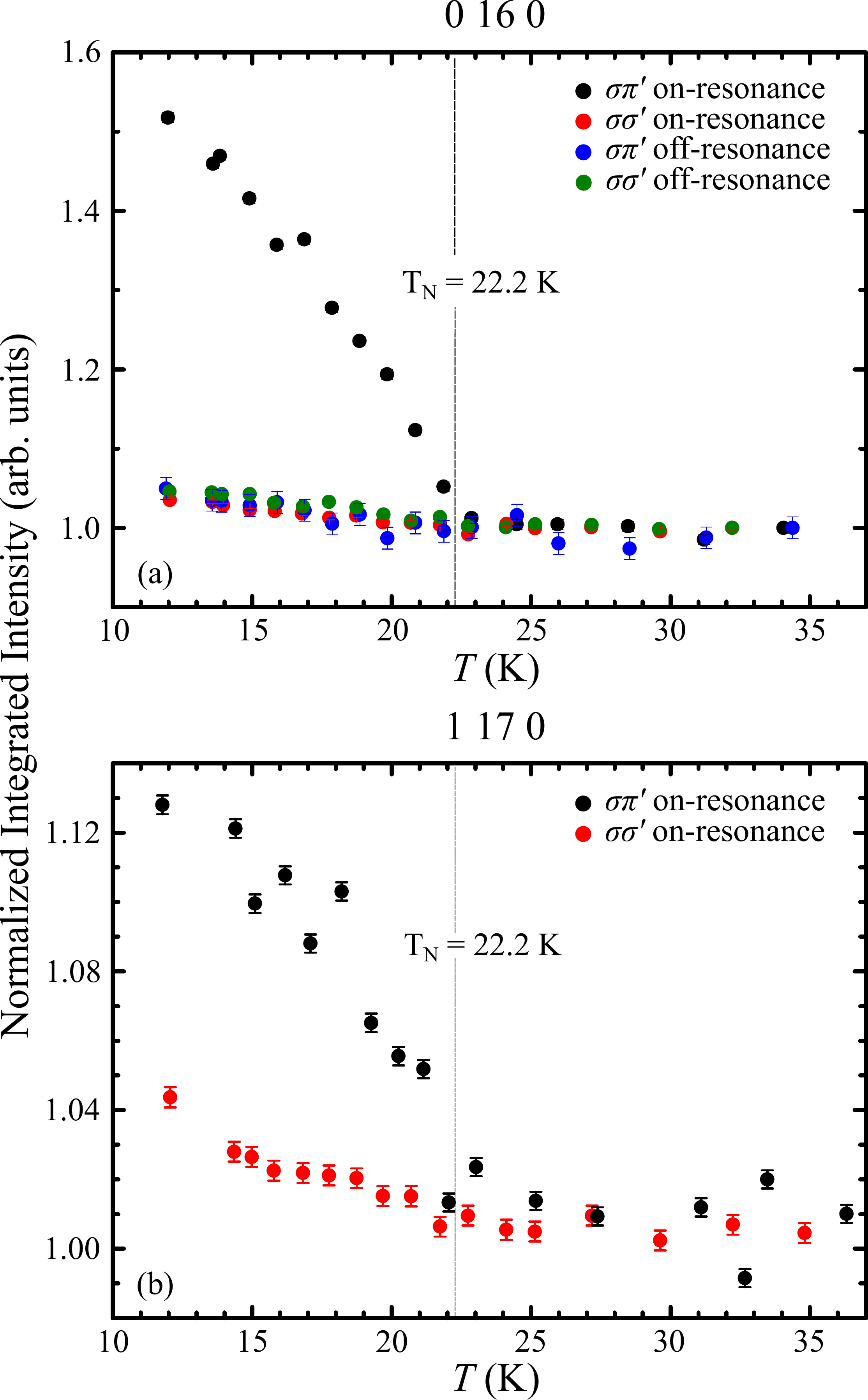}
		\caption{\label{Gd} (a) Temperature-dependence of the integrated intensity of the 0 16 0 Bragg reflection of GdNiSi$_3$ at $\sigma \pi$' and $\sigma \sigma$' polarizations, with x-ray energy either on-resonance at the Gd $L_{II}$ edge ($E=7.924$ keV) or off-resonance ($E=7.885$ keV). (b) Similar to (a) for the 1 17 0 reflection (on-resonance only).}
	\end{figure}
	
	\begin{figure}
		\includegraphics[width=0.48 \textwidth]{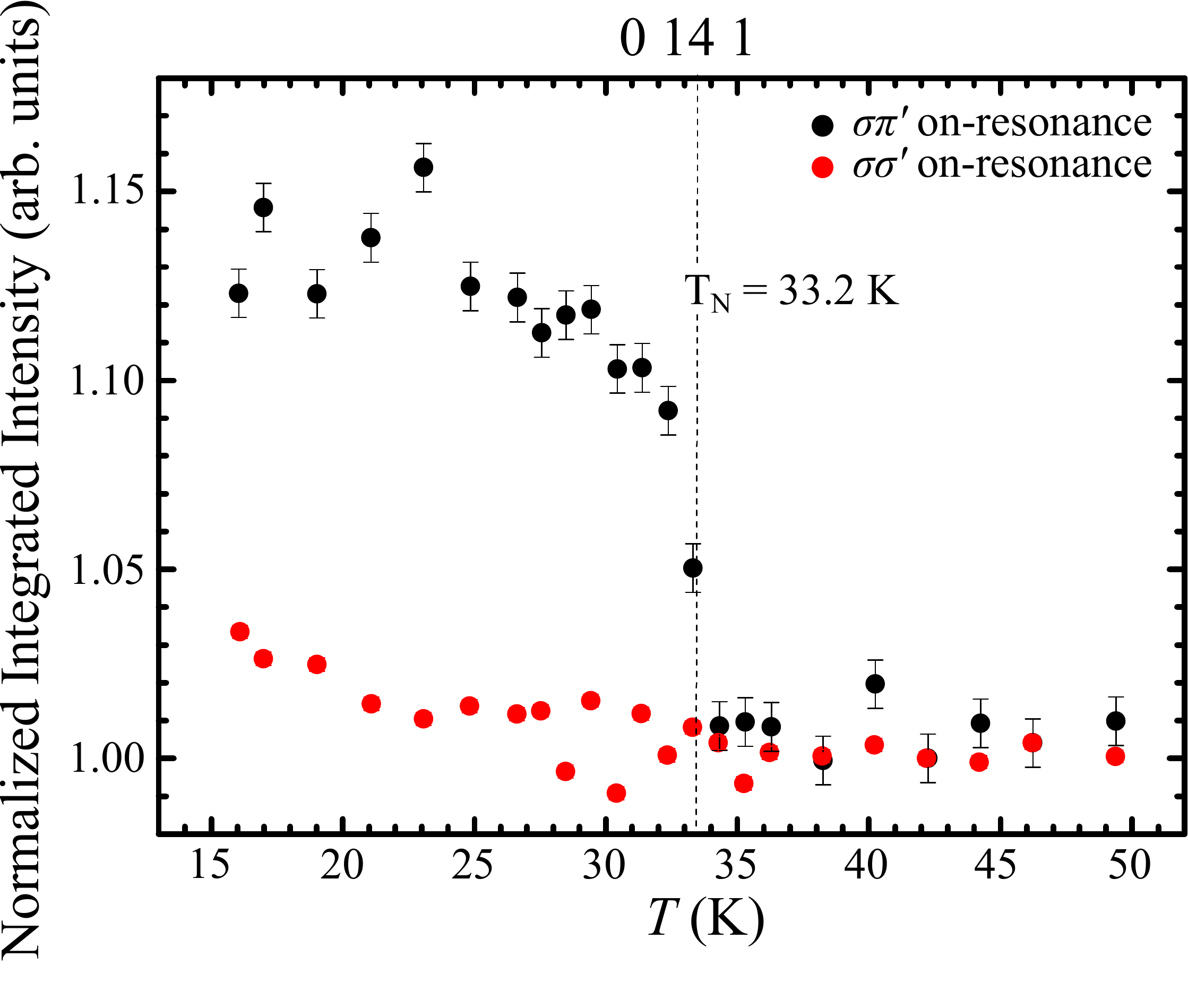}
		\caption{\label{Tb} Temperature-dependence of the integrated intensity of the 0 14 1 Bragg reflection of TbNiSi$_3$ at $\sigma \pi$' and $\sigma \sigma$' polarizations, with x-ray energy either on-resonance at the Tb $L_{II}$ edge ($E = 7.516$ keV).}
	\end{figure}

	\begin{table}[]
		\caption{Observed and calculated intensities of magnetic Bragg reflections for TbNiSi$_3$, normalized by the most intense reflection, using a model with $+-+-$ stacking of ferromagnetic $ac$ planes and magnetic moments parallel either to $\vec{a}$, $\vec{b}$ or $\vec{c}$ directions.} 
		\centering 
		\begin{tabular}{c c c c c c} 
			\hline\hline 
			& & &\multicolumn{3}{c}{+ - + -} \\
			\hline
			(\textit{h k l}) & I$_{obs}$ & & $\vec{m}$ $||$ $\vec{a}$ & $\vec{m}$ $||$ $\vec{b}$ & $\vec{m}$ $||$ $\vec{c}$  \\ [0.5ex] 
			\hline  
			(0 14 1) & 86(7) & & 75 & 61 & 20 \\[0.5ex] 
			(0 16 1) & 40(2) & & 93 & 100 & 25 \\[0.5ex]
			
			(1 19 0) & 98(12) & & 86 & 43 & 45 \\ [0.5ex]
			
			(0 10 0) & 100 & & 100 & 19 & 100 \\[0.5ex] 
			\hline 
		\end{tabular}
		\label{table:comparisson_tb} 
	\end{table}

For Gd- and Tb-based magnetic materials, magnetic x-ray diffraction typically shows strong dipolar resonances at the $L_{2,3}$ edges, with maximum intensities at energies $\sim 2$ eV above the corresponding edge positions \cite{Granado2004,Lora-Serrano2006}. A preliminary x-ray fluorescence scan for GdNiSi$_3$ located the Gd $L_{II}$ absorption edge at 7.922 keV (not shown), and the x-ray energy was subsequently fixed at 7.924 keV to search for magnetic reflections. An extensive search for either commensurate or incommensurate magnetic reflections was then performed, without immediate success. Particularly, no magnetic intensities were observed for integer $hkl$ with $h+k=2n+1$, thereby excluding the magnetic structure of YbNiSi$_3$ \cite{Kobayashi2008} as a possibility for GdNiSi$_3$. 
From this initial survey, one could hypothesize at this point that GdNiSi$_3$ either displays (i) a very complex magnetic structure not covered by our survey scans, with a propagation vector $\vec{k}=[k_x,k_y,k_z]$ where neither of the components are integer or half-integer, or (ii) a particularly simple one ($\vec{k}=[0,0,0]$) preserving the same $C$-centering symmetry element of the space group $Cmmm$ of the chemical structure. 
In the latter case, the reflection conditions for the magnetic and chemical structures would be the same.
	
In order to test hypothesis (ii) above, the intensities of the selected reflections with particularly small structure factors for the chemical structure were collected for different polarizations, with the monochromator and analyzer being both fixed at either $E=7.924$ keV (on-resonance) or $E=7.885$ keV (off-resonance). 
Ideally, the charge sector is expected to show vanishing intensities at pure $\sigma \pi$' polarization, while the magnetic sector would contribute strongly to this channel at dipolar resonances \cite{Hill:sp0084}. 
However, even small polarization leaks of the charge intensities, caused both by small deviations of $2 \theta_{analyzer}$ from 90$^{\circ}$ (see above) and by the horizontal divergence of the beam ($\sim 6$ mrad), are sufficient to compete and in most cases dominate over the magnetic intensities even in $\sigma \pi$' polarization. 
In fact, for $T \gg T_N$, where no magnetic Bragg peaks are present, non-vanishing intensities from polarization leak were observed at $\sigma \pi$' for all investigated reflections in a typical level of $\sim 0.1$\% with respect to the corresponding $\sigma \sigma$' intensities (not shown). 
Fig. \ref{Gd}(a) shows the temperature-dependence of the integrated intensities of the 0 16 0 reflection at $\sigma \pi$' and $\sigma \sigma$' polarizations and both on-resonance and off-resonance, normalized by the corresponding intensities at $T=34$ K ($\gg T_N$). 
The intensities show a smooth temperature-dependence within the studied temperature interval, except for those at $\sigma \pi$' on-resonance, which show a sharp increment below $T_N$. 
Crucially, these conditions for the appearance of extra intensities (polarization, energy and temperature) are exactly those expected for a resonant magnetic x-ray diffraction contribution, providing strong evidence for a simple magnetic structure for GdNiSi$_3$ with propagation vector $\vec{k}=[0,0,0]$. 
The temperature-dependence of an additional reflection 1 17 0 was also studied on-resonance, also showing an enhancement below $T_N$ at $\sigma \pi$' polarization [see Fig. \ref{Gd}(b)].
	
The crystal stucture of $R$NiSi$_3$ with $Cmmm$ space group (Fig. \ref{structure}) shows four $R$ ions per conventional unit cell, i.e., two $R$ ions per primitive cell. 
The presence of magnetic reflections at the same Bragg positions of the charge reflections is consistent with either a parallel or antiparallel alignment of the moments of the two $R$ atoms of the primitive cell. 
A parallel alignment would lead to a ferromagnetic structure, which can be excluded by the macroscopic properties of GdNiSi$_3$ \cite{Arantes2018}. 
An antiparallel alignment of Gd spins in the primitive cell therefore remains as the only plausible solution for the magnetic structure of this material. 
This corresponds to ferromagnetic {\it ac} planes that are stacked in an antiferromagnetic $+-+-$ pattern. 
Concerning the moment direction, a group-theoretical analysis indicated that the $R$ moments must lie along one of the {\bf a}, {\bf b}, and {\bf c}-axes, without canting \cite{Kobayashi2008}. 
The moment direction can be experimentally determined by the magnetic intensities according to $I^{M}(\vec{Q}) \propto \left|\sum_{j}\vec{m}_{j}\cdot\vec{k}_{f}e^{i\vec{\tau}\cdot\vec{r}_{j}}\right|^{2}$ valid for dipolar resonances, where $\vec{Q}$ is the reciprocal-lattice vector for the magnetic Bragg reflection, the sum is over the \textit{j}th resonant magnetic ions in the magnet unit cell, $\vec{m}_{j}$ is the magnetic moment at site \textit{j} and $\vec{k}_{f}$ is the wave vector of the scattered light. 
In the conditions of our experiment for GdNiSi$_3$ and considering absorption corrections \cite{Cullity2001}, this formula leads to the prediction that the magnetic intensity ratio between the 1 17 0 and 0 16 0 reflections would be $r \equiv I^{1,17,0}/I^{0,16,0} = 0.38$ and 1.92 for $\vec{m} || \vec{a}$ and $\vec{m} || \vec{b}$, respectively, while zero magnetic intensity would be expected for both reflections for $\vec{m} || \vec{c}$. 
The observed value $r=0.38(7)$, computed after subtracting the spurious polarization-leaked intensity from the total integrated intensity of each reflection, is thus consistent with $\vec{m} || \vec{a}$, in line with the AFM axis identified by magnetic susceptibility data \cite{Arantes2018}. 
A schematic representation of the magnetic structure identified for GdNiSi$_3$ is displayed in Fig. \ref{structure}(right).

The magnetic structure of TbNiSi$_3$ is determined through the same procedure detailed above for GdNiSi$_3$. 
A fluorescence scan located the Tb $L_{II}$ absorption edge position at $E=7.514$ keV, and the monochromator energy was subsequently set at 7.516 keV. Magnetic intensities were then observed over the same $hkl$ reflections allowed for the chemical crystal structure, revealing that the magnetic structure of TbNiSi$_3$ has the same $+-+-$ stacking pattern of GdNiSi$_3$ (see above). Fig. \ref{Tb} shows the temperature-dependence of the integrated intensity of the 0 14 1 reflection at $\sigma \pi$' and $\sigma \sigma$' polarizations, normalized by the corresponding values at $T=44$ K $\gg T_N$, showing the extra magnetic signal below $T_N$ at $\sigma \pi$'. 
An analysis of the relative magnetic intensities of the 0 14 1, 0 16 1, 1 19 0 and 0 10 0 reflections (see Table \ref{table:comparisson_tb}) show reasonable agreement between experimental and calculated data for $\vec{m} || \vec{a}$, also in line with the AFM axis identified by magnetic susceptibility data \cite{Arantes2018}.
	
	\subsection{Magnetoelastic coupling}
	
	\begin{figure}
		\includegraphics[width=0.48 \textwidth]{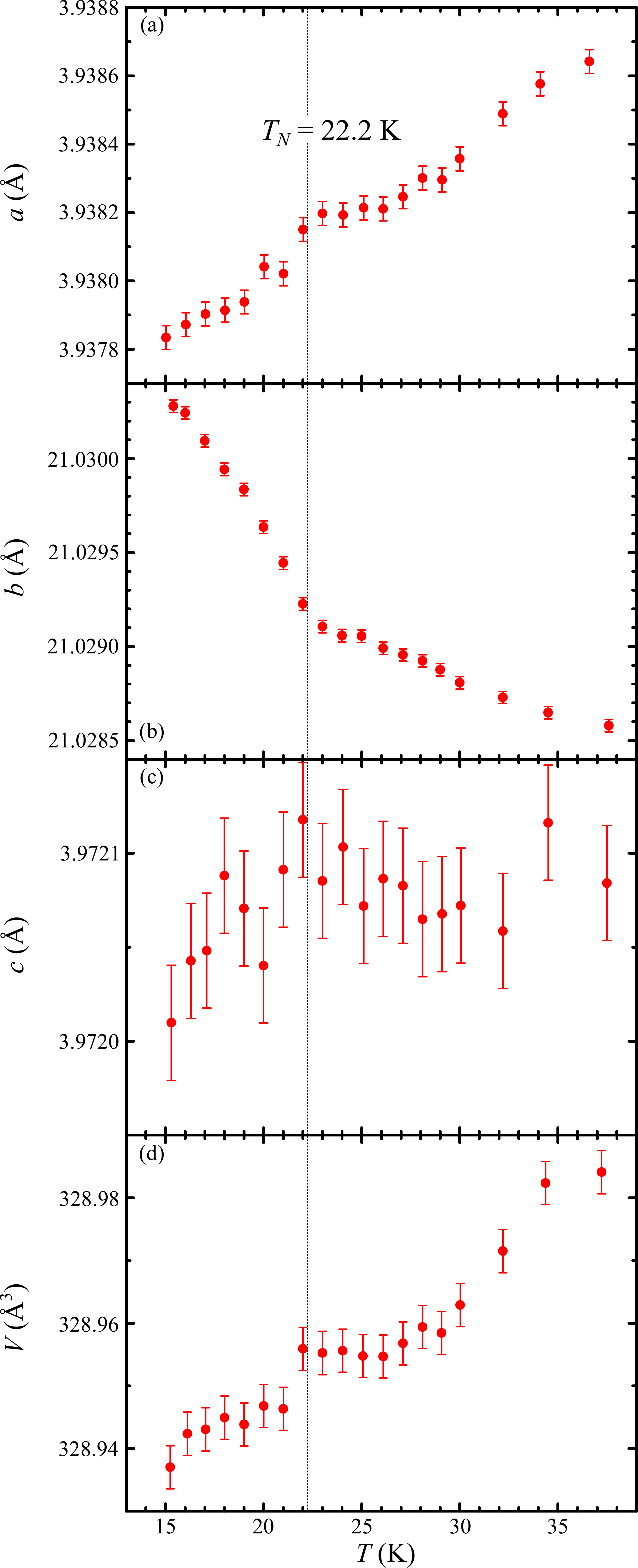}
		\caption{\label{lattpar_gd} Temperature-dependence of $a$, $b$, $c$ orthorhombic lattice parameters and unit-cell volume $V$ for GdNiSi$_3$ (a-d), respectively.}
	\end{figure}

	\begin{figure}
		\includegraphics[width=0.48 \textwidth]{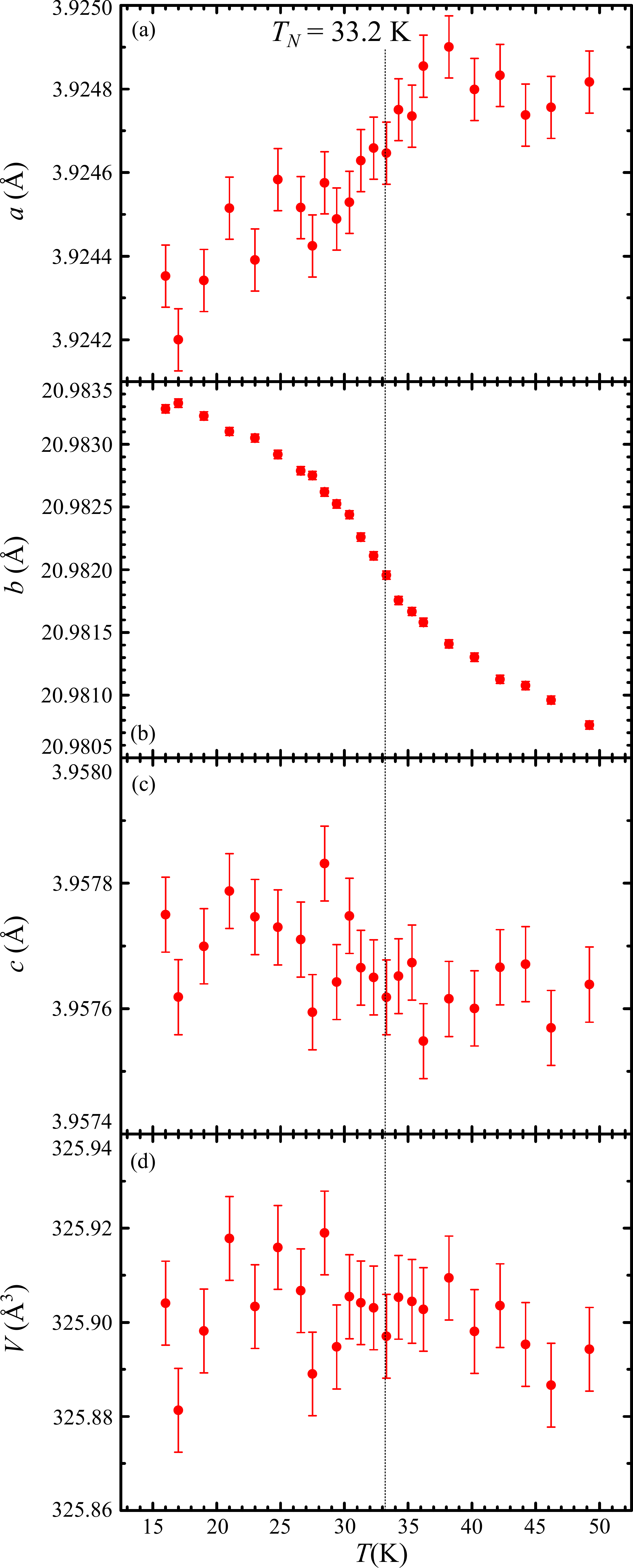}
		\caption{\label{lattpar_tb} Temperature-dependence of $a$, $b$, $c$ orthorhombic lattice parameters and unit-cell volume $V$ for TbNiSi$_3$ (a-d), respectively.}
	\end{figure}
	
The temperature-dependence of the orthorhombic lattice parameters of GdNiSi$_3$ and TbNiSi$_3$ were investigated by an analysis of the angular ($2 \theta$) positions of selected Bragg peaks measured with $\sigma \sigma$' polarization. The results are given in Figs. \ref{lattpar_gd}(a-d) for GdNiSi$_3$ and Figs. \ref{lattpar_tb}(a-d) for TbNiSi$_3$. Both materials show a contraction of $a$ and expansion of $b$ below $T_N$, indicating a magnetoelastic coupling, while $c$ remains approximately constant over the studied temperature interval.
	
	\section{\label{sec:level1}Discussion}
	

The zero-field magnetic structures of GdNiSi$_3$ and TbNiSi$_3$ determined in this work [see Fig. \ref{structure} (right)] are identical. Also, the continuous temperature-dependence of the lattice parameters and magnetic intensities for both materials are indicative of second-order transitions, consistent with previous specific heat measurements \cite{Arantes2018}.
A comparison of the observed magnetic structure with that of YbNiSi$_3$ \cite{Kobayashi2008} [see also Fig. \ref{structure} (left) ] indicates that they have in common the ferromagnetic {\it ac} planes and the antiferromagnetic alignment between such nearest-neighbor planes, indicating that bilayers of $R$ moments are the common magnetic units in this series, at least at zero field. 
On the other hand, these structures differ by the AFM axis (moment directions along $\pm \vec{a}$ for $R$ = Gd and Tb and $\pm \vec{b}$ for $R$ = Yb), and also by the coupling between adjacent bilayers leading to the $+-+-$ and $+--+$ stacking patterns, respectively. 
The change of the magnetic stacking pattern indicates a change of sign of the effective Ruderman-Kittel-Kasuya-Yosida (RKKY) exchange coupling between the second-neighbor $R$ $ac$ layers on reducing the $R$ ionic radius from Gd and Tb to Yb, i.e., on reducing the distance between such layers. In this way, above a certain critical value of $b$, the $+-+-$ pattern is stabilized against the competing $+--+$ ground state of YbNiSi$_3$.
We suggest that the intriguing evolution of the macroscopic magnetic properties of the intermediate members of this series \cite{Arantes2018} is at least in part due to a crossover between the two competing magnetic structures of GdNiSi$_3$ and YbNiSi$_3$. 
We should mention that both magnetic structures differ from the related material CeNiGe$_3$ with a similar crystal structure, where a single crystal neutron diffraction study revealed an incommensurate ground state with propagation vector $\vec{k_2}= [0,0.41,1/2]$ below $T_N=5$ K \cite{Durivault2003}, while a powder neutron diffraction study showed the incommensurate phase coexisting with a commensurate magnetic structure with a $++--$ stacking pattern \cite{Durivault2003}.

The magnetoelastic coupling revealed in our investigation of the orthorhombic lattice parameters [see Figs. \ref{lattpar_gd} and \ref{lattpar_tb} ] also provide insight into the magnetism of this series. As mentioned above, the $+-+-$ magnetic structure for $R=$ Gd and Tb is stabilized by an expansion of $b$ associated with the larger ionic radii of these ions with respect to Yb. The magnetic coupling energy for these compounds is further reduced by an additional expansion of $b$ on cooling below $T_N$, characterizing an exchange striction effect that leads to the observed sign of the magnetoelastic coupling for $R=$ Gd and Tb. It is expected that, for $R=$ Yb, this effect would occur with the opposite sign, since its $+--+$ pattern seems to be stabilized by a contraction of $b$. A detailed thermal expansion investigation of YbNiSi$_3$ is necessary to confirm this scenario. A stabilization of the magnetic structure under lattice expansion is unusual for direct exchange and superexchange coupling mechanisms in insulators, however it is a relatively straightforward phenomenon for metals considering the oscillatory behavior of the exchange integral between local moments mediated by conduction electrons as a function of interatomic distances, such as in the RKKY mechanism. 


	\section{\label{sec:level1}Conclusions}
	
In summary, resonant x-ray diffraction experiments were performed for GdNiSi$_{3}$ and TbNiSi$_{3}$. 
Both compounds have a commensurate magnetic structure with propagation vector $\vec{k}$ = [0 0 0] formed by a $+-+-$ stacking pattern of ferromagnetic $ac$-planes, where Gd and Tb magnetic moments are parallel to $\vec{a}$ axis. 
We also observe a magnetoelastic coupling in both compounds. 
The sign of this coupling along $b$ is consistent with the stabilization of the $+-+-$ stacking under increasing $R$ ionic radius.
A competition between distinct magnetic stacking patterns with similar exchange energies tuned by the size of $R$ sets the stage for the magnetic ground state instability observed along this series.
	
	\begin{acknowledgments}
		
		We thank M. A. Eleot\'erio for technical support at the XDS beamline at LNLS. This work was supported by Fapesp (Grant Nos. 2017/04913-1, 2014/20365-6 and 2011/19924-2) and CAPES.
		
	\end{acknowledgments}

\end{document}